%% file: ieee.tex
\documentclass[conference]{IEEEtran}
\usepackage{cite}
\usepackage{amsmath,amssymb,amsfonts}
\usepackage{algorithmic}
\usepackage{graphicx}
\usepackage{textcomp}
\usepackage{xcolor}
\usepackage{amsmath,amssymb,amsfonts}
\usepackage{textcomp}
\usepackage{graphicx}
\usepackage{times}
\usepackage{microtype}
\usepackage[UKenglish]{babel}
\usepackage{graphicx,dblfloatfix}
\usepackage{blindtext}
\usepackage{mathtools}
\usepackage{multirow}
\usepackage{tabularx}
\usepackage{subcaption}
\usepackage{hhline}
\usepackage{arydshln}
\usepackage{listings}
\usepackage{adjustbox}
\usepackage{color,soul}
\usepackage{booktabs}
\usepackage[margin=1in]{geometry}
\IEEEoverridecommandlockouts

\usepackage{eso-pic}
\begin{document}
\AddToShipoutPictureFG*{\AtPageLowerLeft{\put(\LenToUnit{0.5\paperwidth},\LenToUnit{0.3in}){\makebox[0pt]{\parbox[b]{0.9\paperwidth}{\scriptsize\centering \copyright\,2026 IEEE. Personal use of this material is permitted. Permission from IEEE must be obtained for all other uses, in any current or future media, including reprinting/republishing this material for advertising or promotional purposes, creating new collective works, for resale or redistribution to servers or lists, or reuse of any copyrighted component of this work in other works.\\ Accepted version of the paper published in the 2026 IEEE International Conference on Interdisciplinary Approaches in Technology and Management for Social Innovation (IATMSI). DOI: 10.1109/IATMSI68868.2026.11465435}}}}}

\title{Heuristic-Driven Resource Optimization for Decoupled Memory Channels on FPGAs\\

}

\author{
    \IEEEauthorblockN{Arnab A Purkayastha\IEEEauthorrefmark{1}, Shobhit Aggarwal\IEEEauthorrefmark{2}, Jay Tharwani\IEEEauthorrefmark{3}}
    \IEEEauthorblockA{\IEEEauthorrefmark{1}Western New England University, Member IEEE, Springfield, MA, USA\\
    Email: arnab.purkayastha@wne.edu}
    \IEEEauthorblockA{\IEEEauthorrefmark{2} {Department of ECE}, {The Citadel}, Charleston, NC, USA\\
    shobhit.aggarwal@citadel.edu}
     \IEEEauthorblockA{\IEEEauthorrefmark{3}
    Senior Member IEEE, Morrisville, NC, USA\\
    Email: jtharwan@alumni.uncc.edu}
}

\maketitle

\input{tex/abstract}

\vspace*{0.5 cm}

\input{tex/introduction}
\input{tex/methodology}

\input{tex/setup}

\input{tex/results}

\input{tex/conclusion}

\bibliographystyle{IEEEtran}
\scriptsize
\bibliography{ieee}

\end{document}

%% file: tex/abstract.tex
\begin{abstract}
Determining OpenCL channel depths for Decoupled Access-Execute (DAE) architectures in High-Level Synthesis (HLS) is critical for balancing performance and hardware efficiency. Conventional static analysis tools often over-provision these memory channels to ensure performance safety, leading to significant resource stranding. This paper presents an expansion of an LLVM-based static analysis framework-originally proposed for automated DAE generation, by introducing a hardware-aware refinement layer. We identify a counter-intuitive \emph{Performance Inversion Paradox} linked to baseline memory stalls, where high-stall, grid-based kernels reach peak performance at minimal channel depths and suffer up to a 61\% throughput degradation when depths are increased.
We propose a Heuristic Refinement Layer that optimizes OpenCL channel depths based on this correlation between baseline stalls and performance saturation. Experimental results demonstrate that our expanded framework successfully identifies the \emph{Minimal Peak Performance} point across various benchmarks. By avoiding the performance traps of standard over-provisioning, we achieve a 2.52$\times$ improvement in Resource Efficiency ($\eta$) for an illustrative HotSpot application. This approach maintains peak execution speedup while ensuring the leanest possible hardware profile, outperforming conservative tool-based static predictions.

\end{abstract}

%% file: tex/introduction.tex
\section{Introduction}
\label{sec:Introduction}

The rapid evolution of High-Level Synthesis (HLS) has transformed the landscape of FPGA-based hardware design, enabling developers to abstract complex spatial architectures using high-level languages like C, C++, and OpenCL. By automating the transition from software algorithms to hardware description, HLS has significantly lowered the barrier to entry for accelerating data-intensive workloads in domains such as genomics, finance, and artificial intelligence. However, despite the massive parallel processing capability provided by modern FPGA fabrics, achieving peak theoretical performance remains an elusive goal due to the \emph{memory wall.} As highlighted in architectural studies, the memory wall represents a fundamental performance gap where the data-consumption rate of the internal logic far outpaces the rate at which the memory subsystem can supply that data \cite{elsev}.

In the context of FPGAs, this gap is uniquely challenging; unlike traditional CPUs that rely on deep, fixed cache hierarchies to mitigate latency, FPGAs require designers to manually or semi-automatically orchestrate data movement. When memory-bound applications are ported to FPGAs, the execution pipelines often spend a majority of their operational cycles in a stalled state, waiting for high-latency DRAM transactions to complete. This under-utilization of the FPGA's computational resources negates the energy-efficiency and throughput benefits that hardware acceleration is intended to provide. Consequently, the primary challenge in modern HLS design is not just the implementation of the computational logic, but the engineering of sophisticated memory-interfacing structures that can effectively bridge this wall.

To quantify the impact of the memory wall, we characterized three representative benchmarks on an Intel Arria 10 FPGA. As shown in Table~\ref{tab:baseline_stalls}, the baseline (non-decoupled) versions of these kernels suffer from significant memory-related inactivity. These stall percentages represent a massive under-utilization of the FPGA’s computational resources. For instance, in the HotSpot kernel, the execution engine is idle for over three-quarters of its operational cycles, waiting for memory transactions to complete. This empirical evidence highlights that the primary bottleneck is not the compute logic itself, but the lack of an efficient data-supply mechanism.

\begin{table}[htbp]
\centering
\caption{Baseline Memory Stall Characterization}
\label{tab:baseline_stalls}
\begin{tabularx}{\columnwidth}{@{}lXlc@{}} %
\toprule
\textbf{Benchmark} & \textbf{Domain} & \textbf{Access Pattern} & \textbf{Stalls (\%)} \\ \midrule
HotSpot            & Physics Simulation          & Regular (Grid)          & 76.8\%               \\
BFS                & Graph Analytics             & Irregular (Sat.)        & 40.7\%               \\
RangeK             & Database Search             & Variable (Bursty)       & 3.01\%               \\ \bottomrule
\end{tabularx}
\end{table}

To navigate the constraints of the memory wall, the OpenCL standard for FPGAs provides a specialized abstraction known as \emph{Pipes} or \emph{Channels}. As described in \cite{elsev, Amir:2015}, these channels facilitate high-bandwidth, point-to-point communication between concurrent kernels, allowing data to be streamed directly between logic blocks without the overhead of returning to global memory. This capability is foundational for implementing the DAE framework, a hardware architecture specifically designed to mitigate memory latency.

In a DAE system, the computational kernel is partitioned into two distinct units: the Access unit and the Execute unit. The Access unit is dedicated solely to calculating memory addresses and issuing read/write requests to DRAM, while the Execute unit focuses on the core algorithmic operations. By separating these concerns, the Access unit can \emph{run ahead} of the computation, pre-fetching data into FIFO-based channels. This creates a producer-consumer relationship where the Execute unit pulls data from a ready buffer rather than waiting on high-latency memory transactions. When properly implemented, this decoupling allows the FPGA to overlap memory communication with computation, effectively \emph{hiding} the latency that would otherwise stall the execution pipeline.

While DAE is effective, manual partitioning is complex. Recent work introduced an LLVM-based toolchain to automate this process at the Intermediate Representation (IR) level \cite{elsev}. By analyzing data-flow graphs, this tool automatically identifies predictable access patterns. However, a significant limitation remains: the tool does not give the optimal channel depth. The onus thus lies on the designer/ programmer who generally default to a \emph{one-size-fits-all} channel depth (typically 64) for any complex indexing. This over-provisioning leads to Resource Stranding, where physical Block RAM (BRAM) is occupied but provides no additional performance gain, as many kernels reach \emph{performance saturation} at much lower depths.

This paper extends the foundational LLVM-based DAE work by introducing a Stride Classification logic and a Heuristic-Driven Optimization Layer. We propose that by analyzing the baseline stall characteristics identified in Table~\ref{tab:baseline_stalls}, we can \emph{right-size} channel depths to achieve a pareto-optimal trade-off.

In summary, the key contributions of this work are as follows:
\begin{itemize}
\item \textbf{Expanded LLVM Toolchain:} An enhanced \emph{DAE framework} expanded with a \emph{stride classifier engine}, \emph{heuristic refinement logic}, and full support for \emph{modern opaque pointers}.
\item \textbf{Performance Inversion Discovery:} Identification of a counter-intuitive \emph{Stall-Performance Correlation}, proving that over-provisioning \emph{OpenCL channels} in bandwidth-saturated kernels leads to performance drop.
\item \textbf{Efficiency Optimization:} A refinement methodology metric achieving improvement in Resource Efficiency \emph{right-sizing} channels to their minimal peak-performance depth.
\end{itemize}

 All our source codes are reproducible and freely available to download.\footnote{Repository link provided upon publication.}.

%% file: tex/methodology.tex
\section{Methodology and System Design}
\label{sec:method}

The proposed framework operates as a multi-stage optimization pipeline that transforms high-level OpenCL kernels into resource-optimized DAE architectures. The core of this system is the integration of a front-end LLVM static analysis engine with a back-end heuristic refinement layer. The primary goal is to bridge the gap between \emph{performance-safe} static predictions and \emph{resource-efficient} hardware implementations.

\begin{figure*}[b]
\centering
\includegraphics[width=0.6\textwidth]{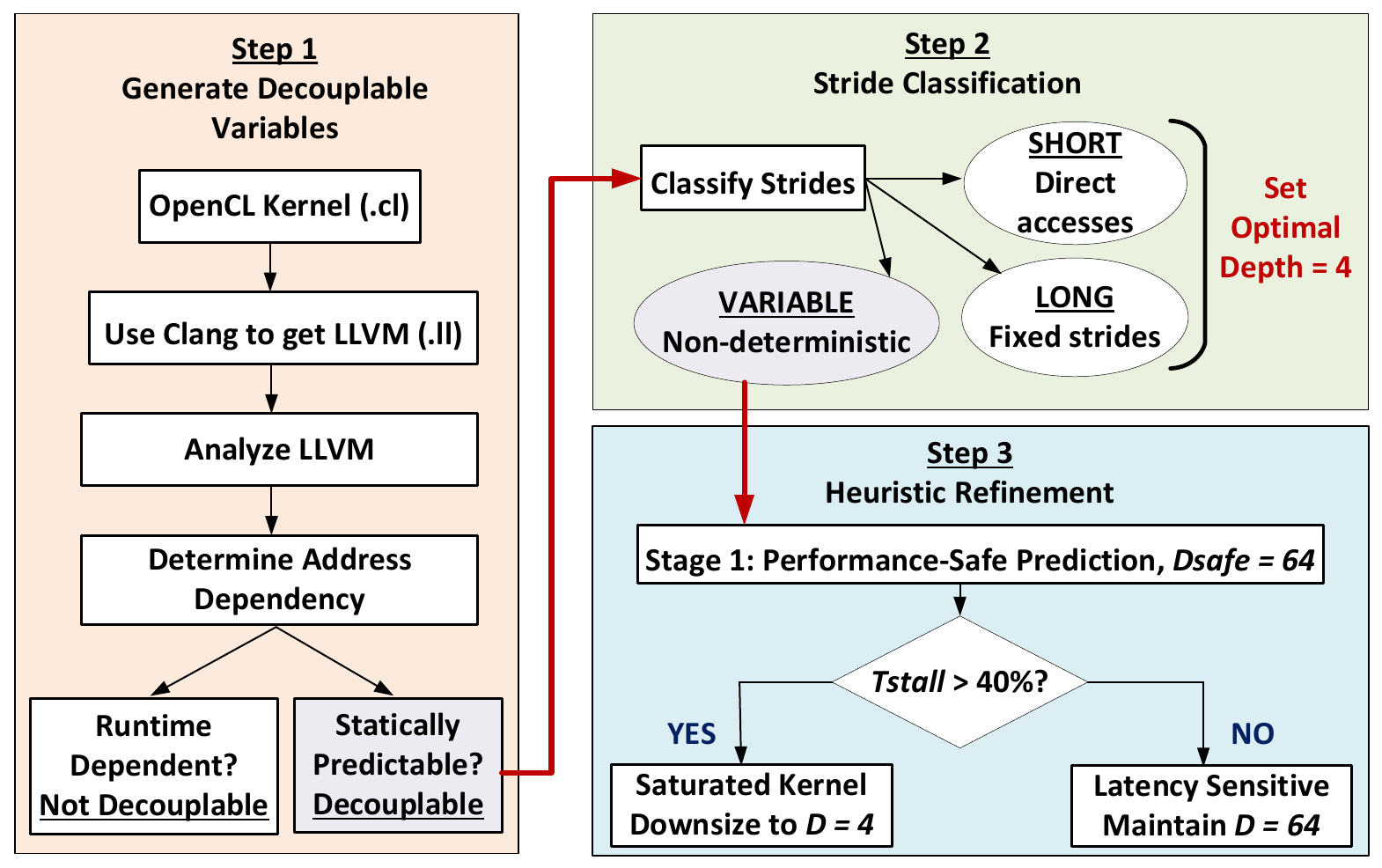}
\caption{Proposed  three-step LLVM optimization tool chain}
\label{fig:toolchain}
\end{figure*}

\subsection{LLVM Static Analysis Engine}
The first stage of our methodology involves an enhanced LLVM-based parser which operates directly on the LLVM Intermediate Representation (IR)\cite{llvm0, llvm}. Parsing at the IR level is critical because it allows the tool to observe the finalized data-flow and memory-addressing logic after initial compiler optimizations.

\subsubsection{Mechanism of the LLVM Parser}
The parser functions by traversing the LLVM Module’s Call Graph and Control Flow Graph (CFG). The analysis follows a three-step instruction-walking process:
\begin{itemize}
\item Iterative Context Identification: The tool first identifies the innermost loops within a kernel. It searches for the Induction Variable (IV), which serves as the primary iterator for memory indexing. By identifying the initial value and the increment step of this iterator, the tool establishes a baseline for calculating address offsets.
\item Instruction Walking (Use-Def Chains): Upon locating a memory access instruction (e.g., a load or store), the parser \emph{walks} backward through the Use-Def Chain. This recursive process analyzes the operands to determine how the effective memory address was calculated. It specifically looks for arithmetic operations—such as multiplications and additions—that link the memory address to the previously identified induction variable.

\item Address Offset Analysis: The address offset analysis focuses on the address calculation instructions. The parser extracts the Base Pointer (the start of the array or buffer) and the Computed Offset. If the offset is found to be a linear function of the induction variable, the tool identifies it as a \emph{Predictable Stride}.

\end{itemize}

\subsubsection{Addressing the Opaque Pointer Challenge}
While earlier versions of the framework relied on explicit LLVM typed pointers, the new version implements a Type Inference engine that inspects instruction metadata and bit-widths to support the  modern Opaque Pointer standard. This ensures accurate stride normalization and access pattern detection across diverse data types, maintaining analytical robustness where previous pointer-dependent methods would fail.

\subsubsection{Stride Classification Logic}
The engine calculates the memory stride ($S$) by evaluating the coefficient of the induction variable within the address calculation logic. We classify memory access into three primary categories:

\begin{itemize}
\item SHORT ($S \leq 2$): Direct or adjacent accesses (e.g., streaming data).
\item LONG ($2 < S \leq 16$): Fixed-distance strides, typical of structured array offsets or small stencils.
\item VARIABLE ($S > 16$ or non-deterministic): Access patterns where the index depends on another memory load (indirect addressing) or complex polynomial math (stencil grids).

\end{itemize}

The stride calculation for a memory instruction $M$ indexed by induction variable $i$ is defined by the following relationship:

\begin{equation}
S = \frac{\Delta \text{Address}}{\Delta i \times \text{sizeof}(\text{Type})}
\label{eq:stride}
\end{equation}

If the parser cannot resolve $S$ to a constant integer during compile-time (static) analysis, it defaults the classification to VARIABLE, triggering the performance-safe buffering logic in the subsequent stage.

\subsubsection{Heuristic Refinement Logic}

The refinement logic operates in two distinct stages to transform the raw parser output into a final hardware configuration.

\emph{Stage 1: Performance-Safe Prediction:}
In the initial stage, the tool prioritizes performance safety. For any access pattern classified as VARIABLE, the engine assumes a non-deterministic memory behavior that could lead to significant pipeline stalls. To mitigate this, it defaults to a conservative maximum channel depth ($D_{max}$):$$D_{safe} = \begin{cases} 4 & \text{if } S \in \{\text{SHORT, LONG}\} \\ 64 & \text{if } S = \text{VARIABLE} \end{cases}$$While $D=64$ ensures sufficient buffering to hide most DRAM latencies, it often results in over-provisioning for kernels that hit bandwidth limits early.

\emph{Stage 2: Stall-Aware Heuristic:}
The proposed extension introduces a \emph{Resource-Aware} check. We observe that kernels with high baseline stall rates ($T_{stall}$) behave as \emph{saturated} systems. For these kernels, adding buffer depth beyond the initial saturation point yields diminishing returns on speedup ($R$). We apply the Stall-Performance Heuristic to refine the depth ($D_{final}$):$$D_{final} = \begin{cases} D_{safe} & \text{if } T_{stall} < \text{Threshold} \\ \text{Saturation\_Point} & \text{if } T_{stall} \geq \text{Threshold} \end{cases}$$ For our target FPGA, we empirically determined the $\text{Threshold} = 40\%$ and the $\text{Saturation\_Point} \in \{4, 16\}$.

\subsection{Integrated Optimization Toolchain}

The proposed framework, illustrated in Figure \ref{fig:toolchain}, operates as an end-to-end pipeline that bridges high-level kernel descriptions with resource-optimized hardware bitstreams. The toolchain transitions through three distinct phases:

\begin{itemize}
\item \textbf{Generate Decouplable Variables (Step 1)}: The process initiates with the conversion of OpenCL kernels into LLVM IR. The engine performs instruction walking to identify decouplable variables by isolating those with statically predictable address dependencies from those with runtime-dependent indexing \cite{elsev}.

\item \textbf{Stride Classification (Step 2)}: Decouplable variables are categorized based on their access strides. While SHORT and LONG strides are directly mapped to minimal depth configurations, VARIABLE strides are flagged for further refinement.

\item \textbf{Heuristic Refinement (Step 3)}: This stage resolves the \textit{Performance-Safe} vs. \textit{Resource-Efficient} conflicts. By incorporating hardware-level stall characteristics ($T_{stall}$), the framework distinguishes between Saturated Kernels (requiring minimal depth) and Latency Sensitive Kernels (requiring deeper buffers), effectively eliminating resource stranding while maintaining peak speedups.

\end{itemize}

%% file: tex/setup.tex
\section{Experimental Setup}
\label{sec:setup}

This section details the hardware and software environments used to validate the proposed heuristic-driven optimization framework. The methodology relies on real-time hardware profiling to bridge the gap between static LLVM IR analysis and physical resource utilization.

\subsection {Hardware and Software Environment}

All experiments were conducted on an Intel Arria 10 GX FPGA. The development environment utilized the Intel FPGA SDK for OpenCL and the Quartus Prime Pro Edition toolchain \cite{khronos:2008,intelOpenCl}.

\subsection{Profiling and Characterization}

To obtain the stall data ($T_{stall}$) required for Step 3 of the toolchain, we utilized the Intel FPGA Dynamic Profiler. Kernels were initially compiled without OpenCL Pipes to identify the raw memory bottleneck. We specifically monitored Load/Store Stalls, which represent the percentage of cycles the execution pipeline is inactive due to memory back pressure. Post-synthesis reports were used to extract Memory Blocks (Block RAM, BRAM), Logic Utilization (Adaptive Logic Module, ALM), and Registers (Flip-Flop, FF) usage for each depth configuration.

\subsection{Benchmarks}

We selected three representative applications from the Rodinia Benchmark Suite \cite{rodinia, rodinia2}. These benchmarks were chosen specifically because they represent distinct memory access behaviors that test the limits of automated HLS depth allocation:

\begin{itemize}
    \item HotSpot (Physics Simulation): Chosen for its regular, grid-based access pattern that paradoxically suffers from high baseline stalls (76.8\%). It serves as the primary test case for Saturated Kernel optimization.
    \item BFS (Graph Analytics): Selected to represent irregular, non-deterministic access patterns that are heavily bandwidth-limited (40.7\% stalls). This tests the tool’s ability to categorize VARIABLE strides that hit performance saturation early.
    \item RangeK (Database Search): Included as a control for Latency-Sensitive applications. With very low baseline stalls (3.01\%), it requires deep buffering ($D=64$) to handle bursty transactions without losing performance.
\end{itemize}

%% file: tex/results.tex
\vspace{0.5pt}
\section{Results and Discussion}

\subsection{Performance Analysis and Stall Correlation}

\begin{figure}[htbp]
\centering
\includegraphics[width=0.8\columnwidth]{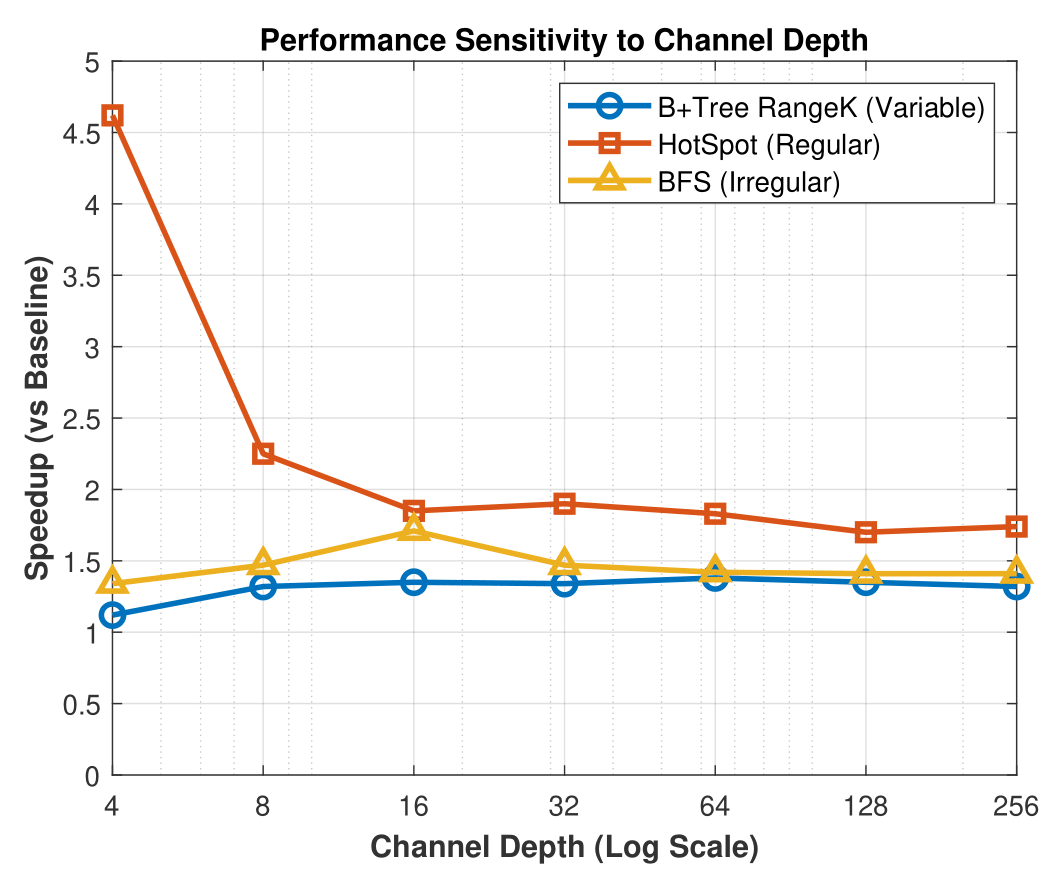}
\caption{Speedup vs. Channel depth for all three applications}
\label{fig:speedup_plot}
\end{figure}

\begin{figure}[htbp]
\centering
\includegraphics[width=0.8\columnwidth]{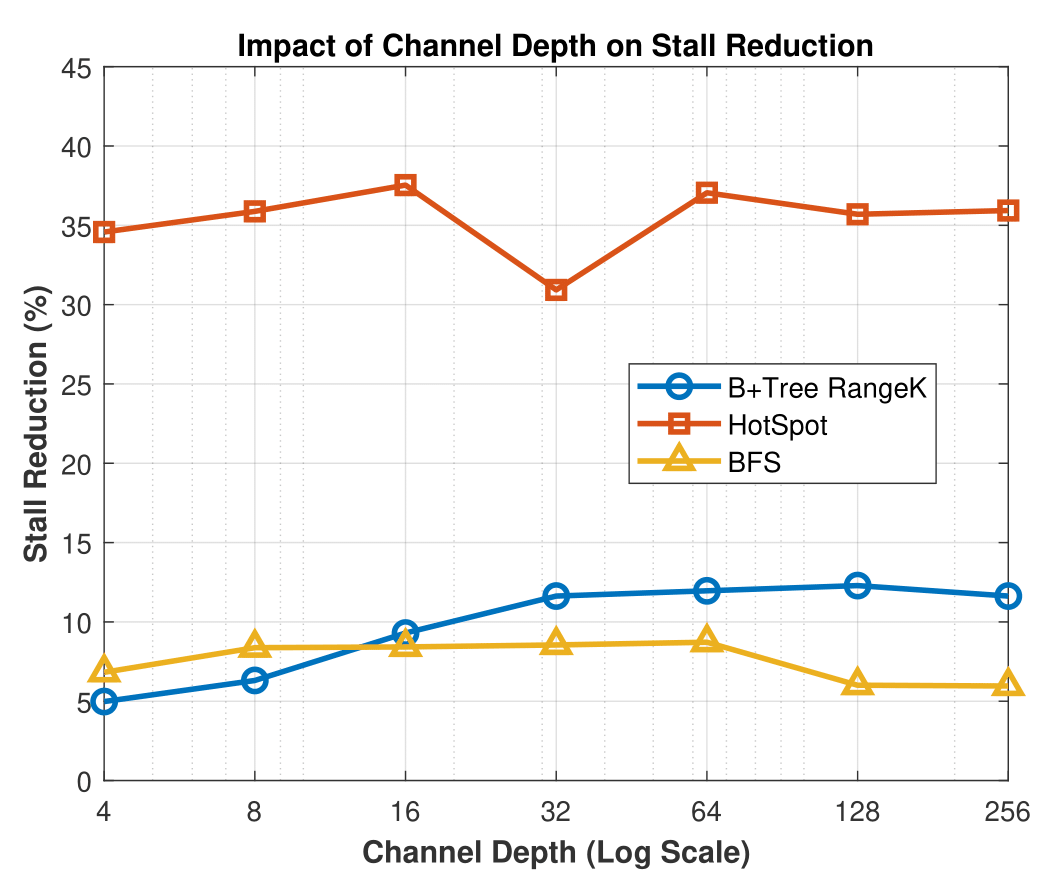}
\caption{DAE optimized stall reduction across channel depths for all three applications}
\label{fig:stalls}
\end{figure}

The efficacy of our heuristic is rooted in the  relationship between baseline memory stall rates ($T_{stall}$) and the resulting speedup across varying OpenCL channel depths. Initial characterization results revealed a significant disparity in memory behavior among the Rodinia benchmarks. For instance, HotSpot and BFS exhibited high baseline stall rates of 76.8\% and 40.7\%, respectively, classifying them as bandwidth-saturated. In contrast, RangeK maintained a much lower stall rate (approximately 3.01\%), indicating a latency-sensitive profile. Figures \ref{fig:speedup_plot} and \ref{fig:stalls} show the Speedup and Stall reduction vs various channel depths (4-256) for the three representative applications.

A critical observation from the results is the \emph{Performance Inversion Paradox}, where increasing hardware resources leads to a drastic reduction in computational throughput.To quantify this loss, we define Performance Degradation ($P_{\Delta}$) as the percentage of peak speedup lost when transitioning to a standard \emph{safe-max} default ($D=64$):

\begin{equation}P_{\Delta} = \frac{S_{peak} - S_{std}}{S_{peak}} \times 100\label{eq:degradation}
\end{equation}

Where, $S_{peak}$ is the maximum achieved speedup and $S_{std}$ is the speedup at the standard channel depth. Applying Equation \ref{eq:degradation} to the HotSpot benchmark, we observe a peak speedup of 4.62$\times$ at a minimal \emph{channel depth of D=4}. However, at $D=64$, the speedup drops to 1.83$\times$, resulting in a 60.3\% performance degradation. Similarly, BFS experiences a degradation of 16.9\% when moving from its optimal depth ($D=16$) to the standard default.

\label{sec:results}

These diminishing returns occur because, in bandwidth-saturated kernels, the execution unit is capped by the DRAM's peak delivery rate. Any additional buffering via deeper \emph{OpenCL channels} only adds management overhead and interconnect pressure without providing further latency-hiding benefits. Conversely, RangeK validates the necessity of the heuristic's second stage; with low baseline stalls, it is highly sensitive to channel depth and achieves its $S_{peak}$ of 1.38$\times$ only at $D=64$.

The non-linear and often contradictory performance trends observed across these benchmarks underscore why choosing a \emph{channel depth} arbitrarily or manually is impractical. Without the automated stall-based classification provided by our expansion, a designer would likely choose a large default depth, unknowingly incurring a significant $P_{\Delta}$ penalty.

\begin{figure*}[htbp]
\centering
\includegraphics[width=0.75\textwidth]{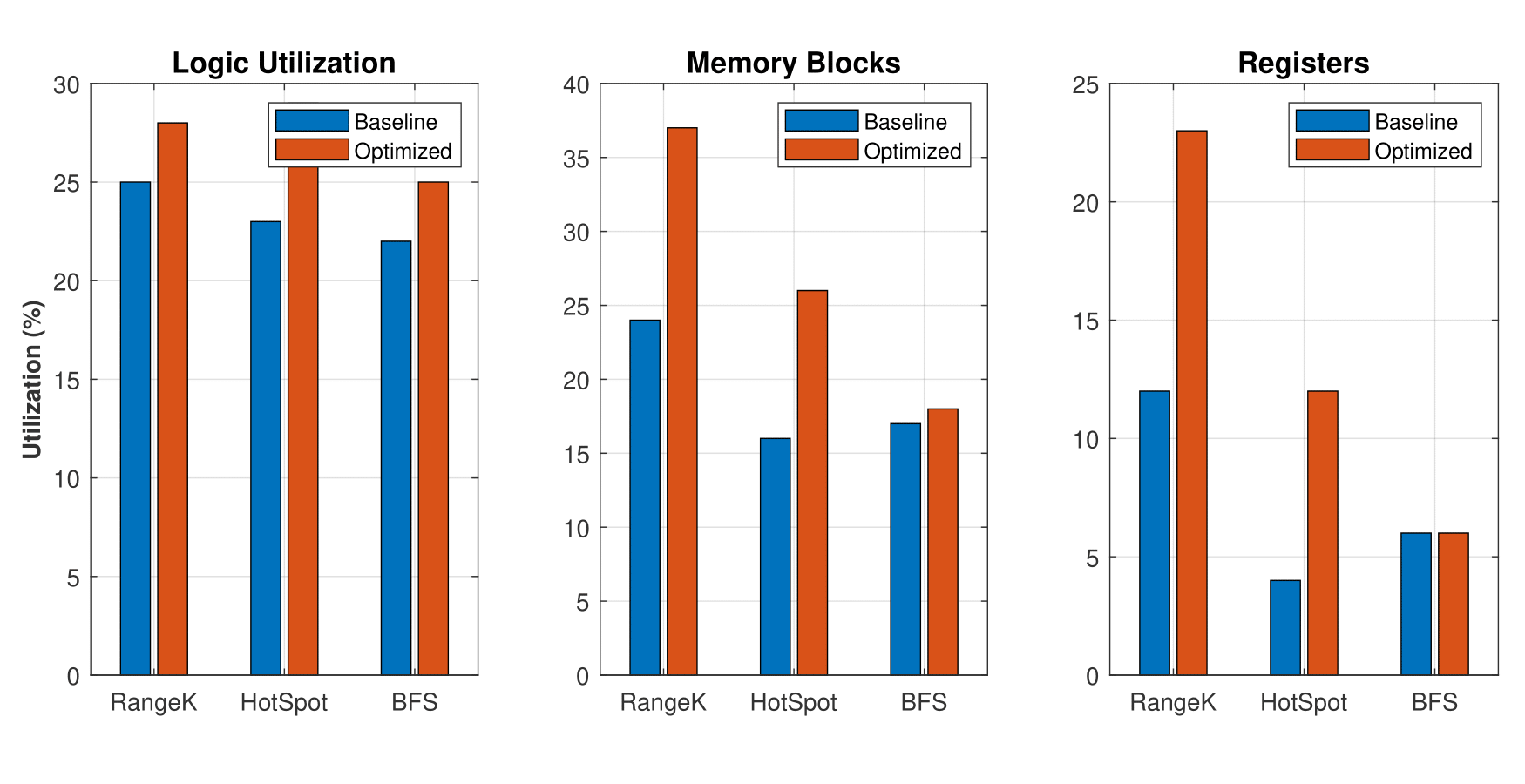}
\caption{Resource utilization comparison between baseline and DAE optimized applications}
\label{fig:resources}
\end{figure*}

\subsection{Resource Overhead Analysis}
The transition from a standard kernel to a DAE framework requires the instantiation of OpenCL channel, introducing a fixed hardware overhead primarily in memory and logic utilization. This base cost is visualized in Figure \ref {fig:resources}, which compares the baseline resource footprint against the optimized decoupled implementations.

As illustrated in the profiling data, adding the DAE hardware results in a measurable resource requirement. For HotSpot, memory block utilization increases from 0.16\% to 0.26\%, while logic utilization rises from 0.23\% to 0.29\%. Similarly, BFS shows a jump in logic utilization from 0.22\% to 0.25\%. However, the data reveals a distinct flatline behavior where the resource footprint remains virtually unchanged as the channel depth scales from \emph{D=4} to \emph{D=64}.

The lack of dynamic resource scaling reported by the profiler is likely due to the underlying hardware mapping on the FPGA fabric. FPGAs utilize discrete, fixed-size memory structures, such as Intel’s M20K blocks. When an OpenCL channel is instantiated, the compiler typically allocates a minimum of one such block to handle the interface. For the relatively small data widths and depth ranges (4 to 64) used in these representative benchmarks, the buffer fits entirely within the initial block allocation. This results in no reported increase in total block count despite the change in channel depth.

It is important to note that while these specific applications demonstrate a flat resource overhead, the complexity of more advanced kernels (incorporating larger data types or significantly more parallel channels) may eventually force the compiler to utilize additional physical blocks at higher depths. Furthermore, even when block counts remain static, deeper channels can increase routing complexity and pressure on the interconnect fabric. This reinforces the need for our heuristic; by selecting the minimal depth required for peak performance, we avoid unnecessary hardware commitment and minimize the risk of routing-induced frequency degradation.

\subsection{Resource Efficiency Analysis}

While the physical resource footprint remains relatively static across the evaluated depth ranges, the \emph{efficiency} with which those resources are utilized varies dramatically. To quantify this relationship, we introduce the Resource Efficiency ($\eta$) metric, defined in Equation \ref{eq:efficiency}. This metric represents the ratio of the achieved speedup to the hardware resource utilization, providing a quantitative measure of the performance return on the hardware \emph{tax} paid for the DAE architecture.

\begin{equation}
\eta = \frac{\text{Speedup}}{\text{Resource Utilization (\%)}} \times 100
\label{eq:efficiency}
\end{equation}

Next, Table \ref{tab:efficiency_metrics} presents a comparison of resource efficiency between the standard \emph{safe-max} default ($D=64$) and the depth selected by our heuristic toolchain.

\begin{table}[htbp]
\centering
\caption{Resource Efficiency ($\eta$) Comparison}
\label{tab:efficiency_metrics}
\resizebox{\columnwidth}{!}{%
\begin{tabular}{@{}llccc@{}}
\toprule
\textbf{Benchmark} & \textbf{Configuration} & \textbf{Speedup} & \textbf{Mem. Utilization} & \textbf{Efficiency ($\eta$)} \\ \midrule
HotSpot  & \textbf{Opt ($D=4$)} & \textbf{4.62$\times$} & 0.26\% & \textbf{17.76} \\
         & Std ($D=64$)         & 1.83$\times$          & 0.26\% & 7.03           \\ \midrule
BFS      & \textbf{Opt ($D=16$)}& \textbf{1.71$\times$} & 0.18\% & \textbf{9.50}  \\
         & Std ($D=64$)         & 1.42$\times$          & 0.18\% & 7.88           \\ \midrule
RangeK   & \textbf{Opt ($D=64$)}& \textbf{1.38$\times$} & 0.13\% & \textbf{10.61} \\ \bottomrule
\end{tabular}%
}
\end{table}

By identifying the performance peak at minimal depths for saturated kernels, the framework significantly maximizes the utility of the committed FPGA fabric.

For the HotSpot benchmark, the heuristic-optimized configuration is 2.52$\times$ more efficient than the standard implementation. As the resource cost is identical for both configurations, selecting the $D=4$ depth is the only method to maximize the performance-per-area return. In contrast, the standard $D=64$ choice represents an \emph{efficiency trap}, where 60\% of the potential performance is lost despite the hardware being fully committed

In a multi-accelerator system where dozens of kernels may be instantiated on a single FPGA, these efficiency gains become multiplicative. By ensuring each kernel operates at its peak $\eta$ value, our toolchain prevents the aggregate performance degradation and resource stranding that occur when multiple kernels are over-provisioned. This automated approach effectively transforms OpenCL channel depth from a manual tuning headache into a systematic optimization parameter, ensuring peak speedup with the leanest possible hardware profile.

%% file: tex/conclusion.tex
\vspace{0.2cm}
\section{Conclusion}
\label{sec:Conclusion}

This paper presented an expansion of our LLVM-based compiler toolchain to automate \emph{OpenCL channel depth} selection. This work addresses a fundamental limitation in automated decoupled architectures: the lack of systematic channel optimization. By correlating static IR analysis with hardware-level memory stall characterization, the framework eliminates suboptimal hardware configurations and \emph{resource stranding}. Our findings reveal a \emph{Performance Inversion Paradox} in bandwidth-saturated kernels, where arbitrary over-provisioning leads to significant throughput loss. We identify these regimes using a 40\% baseline stall threshold, a heuristic marker that classifies a kernel as bandwidth-saturated. By utilizing this threshold, the framework successfully isolates the optimal channel depth required to maintain peak throughput without incurring performance penalties.The impact of this optimization is quantified through the \emph{Resource Efficiency} ($\eta$) metric, which evaluates the performance return on hardware investment. For the \emph{HotSpot} benchmark, the toolchain identified that a minimal channel depth of $D=4$ provides the optimal balance of throughput and area, yielding a 2.52$\times$ efficiency improvement compared to the standard $D=64$ default. Ultimately, this expansion transforms \emph{OpenCL channel depth} from a manual, trial-and-error tuning parameter into a deterministic optimization variable. By ensuring peak efficiency, this framework enables the development of denser and more balanced FPGA accelerators, providing a scalable pathway for complex hardware designs in resource-constrained environments.

\vspace{0.3cm}